\documentclass[final,5p,times,twocolumn]{elsarticle}

\usepackage{amssymb}
\usepackage{amsmath}

\journal{}

\begin{document}

\begin{frontmatter}

\title{Limiting Energy Loss Distributions for Multiphoton Channeling Radiation}

\author{M.V.~Bondarenco}
\ead{bon@kipt.kharkov.ua}
\address{%
NSC Kharkov Institute of Physics and Technology, 1 Academic St., 61108
Kharkov, Ukraine
}%

\date{\today}

\begin{abstract}
Recent results in the theory of multiphoton spectra for coherent
radiation sources are overviewed, with the emphasis on channeling
radiation. For the latter case, the importance of the order of
resummation and averaging is illustrated. Limiting shapes of
multiphoton spectra at high intensity are discussed for different
channeling regimes. In some spectral regions, there emerges an
approximate correspondence between the radiative energy loss and the
electron integrals of motion.

\end{abstract}


\end{frontmatter}


\section{Introduction}

Ultrarelativistic electrons moving periodically in coherent
radiation sources produce fairly monochromatic radiation. The
intensity of this radiation is proportional to the length of the
periodic or confining structure, and if it is sufficiently high, the
electron is likely to emit several photons during its passage. The
problem of multiple photon emission per electron plays significant
role for crystalline radiators with thickness $\gtrsim1\% X_0$
\cite{Bak}, Compton sources with optical laser pulse energy
densities
$\gtrsim1\frac{\text{J}}{\mu\text{m}^2}\frac{E_e}{\text{GeV}}$
\cite{Telnov,DiPiazza}, synchrotron radiation \cite{AccelHandbook},
undulators with the number of periods $\gtrsim 10^2$
\cite{FEL-multiphot}. At that, multiple photon emissions from
different electrons can be discriminated by means of coincidences
between detectors, except in cases when the electron beam density is
very high (e.g., in SASE FELs), and so only the problem of photon
pileup from one electron remains.

Physically, the multiple photon emission from an electron in a
strong electromagnetic field may be viewed as a kind of
electromagnetic showering. It greatly simplifies if, as is
typical for many coherent radiation sources, energies of emitted photons are much smaller than the electron energy:
\begin{equation}\label{omega-ll-Ee}
\omega\ll E_{e}.
\end{equation}
Then, in the first approximation, it is permissible to count only
photons stemming directly from the \emph{classical} current
generated by the initial electron, and neglect $e^+ e^-$ pairs
produced by those photons downstream.

For different experiments, though, the quantities of interest under
conditions of multiple photon emission may differ. For instance, in
synchrotrons one is concerned with lag fluctuations of an electron
accelerated by the running wave, in FELs -- electron bunching, in
Compton sources -- electron energy straggling before its last
interaction with a photon yielding the observed gamma-quantum. For
coherent radiation in oriented crystals, one does not really care
about the spent electron, but instead, measurements of the radiation
are usually carried out with calorimeters, and are restricted to spectra
integrated over (relativistically small) emission angles. Then, only
the aggregate energy of all the nearly forward moving photons from
one electron is measured. In the present article, we will discuss
the latter case.

\section{Resummation technique and its alternatives}

As was mentioned above, the possibility of photon pileup in the
calorimeter causes significant deviation of the multiphoton
probability (i.e., the radiative energy loss) spectrum $dw/d\omega$
from the single-photon spectrum
$\frac{dw_1}{d\omega_1}=\frac1{\omega_1}\frac{dI_{\text{cl}}}{d\omega_1}$
predicted by classical electrodynamics, in spite of condition
(\ref{omega-ll-Ee}). Nonetheless, a multiphoton spectrum can be
deduced from the corresponding single-photon one, as is described
below.

Technically, the situation is the  simplest when one's knowledge of
initial conditions for the electron and the target is sufficient for
completely specifying the electron's trajectory. Then, presuming
that photons exert negligible back action on the electron, their
probability distribution obeys Poisson statistics
\cite{Glauber,BK}\footnote{Deviations from Poisson statistics may
occur, e.g., in FELs, when a strong optical field seeded in the
cavity stimulates emissions at the same frequency. But for gamma-ray
sources, the transparency of any medium is too high to confine the
photons and attain appreciable photon occupation numbers.}
\begin{equation}\label{Poisson}
\frac{dW_n}{d\omega_1\ldots
d\omega_n}=W_0\frac1{n!}\frac{dw_1}{d\omega_1}\ldots\frac{dw_1}{d\omega_n},
\end{equation}
where $n$ is the number of emitted photons,
$W_0=e^{-\int_0^{\infty}d\omega_1\frac{dw_1}{d\omega_1}}$ is the
photon non-emission probability, and factor $1/n!$ accounts for
photon equivalence. It is important to note, though, that the
statistical independence of photon emission events does not imply
that they occur sequentially in time. For instance, in a coherent
state of the quantized electromagnetic field (a closest counterpart
to a classical electromagnetic field) \cite{Glauber}, there is a
quantum superposition rather than mere statistical mixture of photon
number states, which is indicative of their simultaneous generation.

To derive from (\ref{Poisson}) the radiation spectrum measured by a
calorimeter, one must integrate the fully differential spectrum
for any definite number of emitted photons over their entire phase
space, under the restriction of the sum of photon energies to a
certain value $\omega$, and then sum over photon multiplicities:
\begin{equation}\label{generic-sum-int}
\frac{dw}{d\omega}=\sum_{n=1}^\infty \int d\omega_1\ldots d\omega_n
\frac{dW_n}{d\omega_1\ldots d\omega_n}
\delta\left(\omega-\sum_{k=1}^n\omega_k\right).
\end{equation}
Series (\ref{generic-sum-int}) can further be resummed\footnote{The
term ``resummation" was coined in the theory of radiative
corrections at high-energy particle scattering (see, e.g.,
\cite{EW-boson-resum}). In view of the apparent technical
similarity, it seems to be proper also in the field of
coherent radiation sources.} to all orders, by Laplace
transforming it, whereafter it exponentiates, and then can be
transformed back to the $\omega$ variable:
\begin{equation}\label{generic-eq}
\frac{dw}{d\omega}=\frac1{2\pi i}\int_{c-i\infty}^{c+i\infty}ds e^{s\omega+\int_0^{\infty} d\omega_1 \frac{dw_1}{d\omega_1}\left(e^{-s\omega_1}-1\right)}
-W_0\delta(\omega).
\end{equation}
The integration contour in (\ref{generic-eq}) is flexible in the
complex $s$ plane, but the integrand is not the only possible one,
too: The upper limit of the integral in the exponent can be actually
set equal to any value greater than $\omega$, which can be expedient
when dealing with incoherent radiation component
\cite{Bondarenco-paper}. For single-photon spectra with a relatively
sharp upper end $\omega_0\ll E_e$, it is usually sufficient to let
the upper integration limit equal to $\omega_0$.

The practical benefit of the exponentiation is that it reduces
the number of integrals to be handled. If one attempts instead to
calculate termwise the probabilities of emission of a certain number
of photons via Eg.~(\ref{generic-sum-int}), computation of the
multiple integrals even numerically quickly becomes prohibitive.
Sometimes, nonetheless, such multiple integrals are computed by
Monte-Carlo -- see, e.g., Ref. 2 of \cite{DiPiazza}.

An approach alternative to (\ref{generic-eq}) is to view
$dw/d\omega$ as a solution of an integro-differential equation with
the single-photon spectrum playing the role of its kernel.
Specifically, presuming that $dw_1/d\omega_1$ is proportional to the
target thickness $L$, one can formally derive the standard transport
equation
\begin{equation}\label{kinetic-eq1}
\frac{\partial}{\partial L}\Pi(E_e-\omega)=\int_0^{\infty} d\omega_1 \frac{\partial}{\partial L}\frac{dw_1}{d\omega_1}\left[\Pi(E_e-\omega+\omega_1)-\Pi(E_e-\omega)\right]
\end{equation}
holding for the electron energy distribution function
\begin{equation}\label{def-Pie}
\Pi(E_e-\omega)=\frac{dw}{d\omega}+W_0(L)\delta(\omega).
\end{equation}
Equation (\ref{kinetic-eq1}) might also serve as an alternative
justification for the $1/n!$ factor in (\ref{Poisson}), by virtue of
the identity
\begin{eqnarray*}
\frac{\partial}{\partial L}\int d\omega_1\ldots d\omega_n \prod_{k=0}^{n}\frac{dw_1}{d\omega_k}\delta\left(\omega-\sum_{k=1}^n\omega_k\right)\qquad\qquad\nonumber\\
=n\int d\omega_1\ldots d\omega_n \prod_{k=0}^{n-1}\frac{dw_1}{d\omega_k}\frac{\partial}{\partial L}\frac{dw_1}{d\omega_n}
\delta\left(\omega-\sum_{k=1}^n\omega_k\right),
\end{eqnarray*}
should it not be the fact that each photon may actually be
coherently generated along a large length, simultaneously with
formation of other photons.

Another popular approach for computing multiphoton emission
probabilities during semi-classical electron passage is to divide
the trajectory into several pieces, the interference of radiation
from which is expected to be negligible, and define a probabilistic
process of photon emission or non-emission on the junctions. It also
permits taking into account incoherent scattering during radiation,
and other accompanying processes. That approach led to construction
of Monte Carlo simulators widely used in modern practice
\cite{Bak,BKS-UFN,Artru-code,Guidi-Bandiera-Tikhomirov}. At the
present state of the art, such generators usually employ kernels of
generation of only one photon, which is subsequently iterated -- as
well as in the resummation approach described above.

\begin{figure}
\includegraphics[width=85mm]{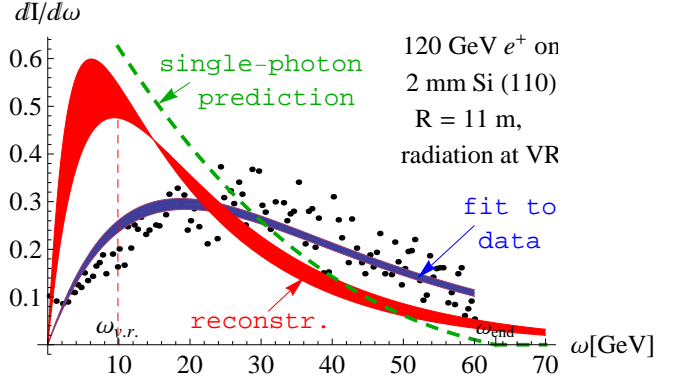}
\caption{\label{fig:Compar-Lietti} Simplified reconstruction of the
single-photon spectrum from multiphoton one, for radiation at volume
reflection. Points, data of \cite{Lietti} for a 2 mm crystal bent to
radius 11 m. Blue band, fit of $\omega\frac{dw}{d\omega}=A\omega
e^{-\alpha\omega}$ to experimental data, the adjusted parameters
being $A=0.042\pm0.003\text{ GeV}^{-1}$ and
$\alpha=0.053\pm0.002\text{ GeV}^{-1}$. Red band, the reconstructed
single-photon spectrum
$\omega_1\frac{dw_1}{d\omega_1}=e^{-\alpha\omega_1}-e^{-\frac{\alpha}{1-A/\alpha}\omega_1}$.
Dashed green curve, prediction for the single-photon spectrum of
coherent bremsstrahlung in a bent crystal \cite{Bond-CBBC}.}
\end{figure}


Adopting an analogy with high-energy physics, one can envisage
further development of simulators to include non-factorized
evaluation of two-photon emission along with the appropriate account
of virtual radiative corrections, simulation of a shower including
coherent photon emission from $e^+e^-$ pairs, etc. But for
qualitative analysis, the resummation method within the region of
its validity is advantageous by its simplicity. Below, we will
confine ourselves to discussing a few examples of its application.

Before we proceed, it is worth pointing out that in case when there
are no random parameters on which the spectrum significantly depends
and which are to be averaged over, relation (\ref{generic-eq}) can
be \emph{inverted} to reconstruct the single-photon spectrum from
the observed multiphoton one \cite{Bondarenco-paper}. In
Fig.~\ref{fig:Compar-Lietti} we show an example where a
single-photon spectrum of radiation at volume reflection was
reconstructed (in a simplified fashion) from the experimental data.
Without taking multiphoton effects into account, the agreement
between the experiment and the theory leaves much to be desired
\cite{Bondarenco-RREPS2010}, but after the reconstruction it becomes
acceptable. Needless to say, the reconstruction procedure is expected
to be efficient only if the radiation intensity (photon
multiplicity) is not too high. In what follows, we will discuss
cases of high intensity and show that the corresponding spectra may
have diverse shapes depending on the physical conditions.

\section{High-intensity limit and the departures from Gaussianity}

For pure coherent radiation, typically characterized by an edgy
single-photon spectrum, multiple photon emission imitates the
appearance of higher spectral harmonics. With the increase of the
radiation intensity, such quasi-harmonics proliferate and coalesce
into universal forms depending on just few parameters. For a
definite and smooth electron trajectory\footnote{Such conditions can
be rather well met in undulator radiation or Compton sources.}, the
(central) limiting form of the multiphoton spectrum is a Gaussian,
as one would expect:\footnote{In \cite{Bondarenco-paper}, moments
$\overline{\omega_{1}}=\int_0^{\infty}d\omega_1\omega_1\frac{dw_\text{1c}}{d\omega_1}$
and
$\overline{\omega_{1}^2}=\int_0^{\infty}d\omega_1\omega_1^2\frac{dw_\text{1c}}{d\omega_1}$
for the coherent radiation component
$\frac{dw_\text{1c}}{d\omega_1}$ were designated as
$\overline{\omega_{1}}_{\text{c}}$ and
$\overline{\omega_{1}^2}_{\text{c}}$. Here, for simplicity we will
not be considering the incoherent contribution at all, therefore
omitting subscript c.}
\begin{equation}\label{Gauss}
\frac{dw_\text{c}}{d\omega}\simeq \frac{1}{\sqrt{2\pi
\overline{\omega_{1}^2
}}}\exp\left[-\frac{(\omega-\overline{\omega_{1}
})^2}{2\overline{\omega_{1}^2 }}\right].
\end{equation}
The corresponding regime may be physically categorized as a normal
random walk. At large $\omega$ beyond the maximum, the spectrum
decreases closer to an ordinary exponential rather than Gaussian law
\cite{Bondarenco-paper}. That can be interpreted in terms of a
maximal-step walk, at which the probability is a product of
probabilities for each maximal step, the number of steps being
$\sim{\omega/\omega_0}$.

A correction to (\ref{Gauss}) breaking the scaling was obtained in
\cite{Bondarenco-paper}. It engages the third spectral moment
related to skewedness. Besides rendering the asymmetry to the spectrum, it
slightly redshifts its maximum.

Another source of departure from the Gaussianity, even in the
absence of dependence of the underlying single-photon spectrum on
external parameters, is the possible infinity of the spectral
moments defining the Gaussian. For instance, in the case of coherent
bremsstrahlung (at a highly over-barrier electron motion), the
dependence on electron impact parameters is negligible, but yet there is
an incoherent radiation component described by Bethe-Heitler
single-photon spectrum $\frac{dw_1}{d\omega_1}=\frac{a}{\omega_1}$,
for which the spectral moments diverge in the ultraviolet (or more
precisely, are dominated by the upper endpoint $\omega\to E_e$,
around which our soft-photon resummation approach breaks down). That
leads to a modification of the limiting spectrum shape by a
power-law tail persisting at high $\omega$. The tail, however, does
not override the whole spectrum, whose maximum is still mainly
determined by the coherent part of the spectrum and its convergent
first moments. Therefore, the diffusion regime may be regarded as
\emph{weakly} anomalous, and the resulting distribution as
intermediate between Gaussian and L\`{e}vy \cite{Bondarenco-paper}.
Within our framework, the anomaly must keep weak, i.e. its index
$a=L/X_0$ must remain small, or else the shower will not be
dominated by soft photons.

\section{Channeling radiation: the impact of initial state
averaging}\label{sec:ChR-averaging}

Our formulation so far held for radiation from a completely
prescribed electromagnetic current. Next, it is important to ask what happens in
case of channeling radiation, when the radiation intensity, and
therewith the photon multiplicity, strongly depends on the
electron/positron oscillation amplitude in the channel, in turn
depending on the electron impact parameter and the angle of entrance
to the crystal. Clearly, in the latter case the multiphoton spectrum
must be duly averaged over the initial electron beam.

An important point here is that the averaging must be
carried out \emph{after} the resummation, and this order of
operations is significant under the condition of high photon
multiplicity. Clearly, in Eq. (\ref{generic-eq}), the average of an
exponential does not reduce to an exponentiated average,
\begin{equation}
\left\langle e^{\int_0^{\infty} d\omega_1 \frac{dw_1}{d\omega_1}\left(e^{-s\omega_1}-1\right)}\right\rangle\neq
e^{\int_0^{\infty} d\omega_1 \left\langle \frac{dw_1}{d\omega_1}\right\rangle\left(e^{-s\omega_1}-1\right)}.
\end{equation}
Similarly, in Eq.~(\ref{kinetic-eq1}),
\begin{equation}
\left\langle \Pi(E_e-\omega+\omega_1)\frac{\partial}{\partial
L}\frac{dw_1}{d\omega_1} \right\rangle\neq \left\langle \Pi(E_e-\omega+\omega_1)
\right\rangle \left\langle \frac{\partial}{\partial
L}\frac{dw_1}{d\omega_1} \right\rangle,
\end{equation}
invalidating the transport equation for the averaged distribution
function with the averaged kernel. In fact, such equations were
heuristically used in the literature both for
overbarrier motion and for channeling conditions
. At that, sometimes they involved adjustable parameters to be
determined from fits to observable spectra. However, since the
fraction of particles near the bottom of the channel radiates
weakly, but contributes to the total number of passed electrons
strongly, estimates even of the mean photon multiplicity per
electron based on kinetic equation with an averaged kernel can be deceptive. 

A related issue is that the knowledge of an \emph{averaged}
single-photon spectrum without detailing its dependence on the
electron initial conditions is \emph{insufficient} for deducing the
(averaged) \emph{multi}photon spectrum. Indeed, the averaged
single-photon spectrum $\left\langle \frac{\partial}{\partial
L}\frac{dw_1}{d\omega_1} \right\rangle$ depends on a single variable
$\omega_1$, whereas the averaged multiphoton spectrum nontrivially
depends on two variables: $\omega$ and the target thickness $L$,
while the dependence on $L$ is intertwined with the dependence on
the electron initial conditions. It should be stressed yet that the
dependence of the radiation on initial conditions must include both
the incidence angle and the impact parameter. Whereas the dependence
on the former is measurable in principle (cf., e.g.,
\cite{Bavizhev}), the dependence on the latter is not (there are no
relativistic beams of angstrom transverse size), and thus inevitably
must be modeled theoretically.

Let us now discuss typical high-intensity channeling radiation
spectrum shapes, for different types of confining wells, and for
simplicity neglecting dechanneling.

\subsection{Positron channeling}\label{subsec:positron-ch}

The averaged  multiphoton spectrum is most easily tractable for
channeling of positrons, when the continuous potential of aligned axes or
planes may be regarded as approximately parabolic in the entire channel (as is known to be
the case for silicon crystals in orientation $(110)$,
$\left\langle111\right\rangle$, or $\left\langle100\right\rangle$).
Then, the particle trajectories in the channels are given by
harmonic curves, and the single-photon radiation spectrum is
explicitly expressible, too. It is proportional to the amplitude
squared of the transverse motion, and therefore to the electron
transverse energy $E_{\perp}$, whereas the differential of the phase
space over which the initial conditions are uniformly distributed
can be written as $dE_{\perp}^{\mathcal{D}}$, with $\mathcal{D}$ the
channeling well dimensionality:
\begin{equation}\label{averaging-positron-def}
\left\langle \ldots \right\rangle=\frac1{E_{\perp\max}^{\mathcal{D}}}\int_0^{E_{\perp\max}}dE_{\perp}^{\mathcal{D}} \ldots=\mathcal{D}\int_0^1 d\xi\xi^{\mathcal{D}-1} \ldots .
\end{equation}
Here value $\mathcal{D}=2$ corresponds to capture to axial
channeling from a beam with angular divergence
$>\theta_c$,\footnote{In that case, there arises a practical
question how to disentangle channeling radiation we are concerned
with from that from overbarrier particles. One possibility is to use
slightly bent crystals, in which channeled particles considerably
deflect but do not significantly change their radiation spectrum,
and then select radiation events in coincidence with deflected
particles -- see, e.g., \cite{Lietti}.} $\mathcal{D}=1$ -- to
capture to planar channeling from a beam with angular divergence
$>\theta_c$, or to axial channeling from a beam co-aligned with the
axis and having divergence $\ll\theta_c$, and $\mathcal{D}=1/2$
corresponds to planar channeling from a perfectly aligned beam with
divergence $\ll\theta_c$.


The analysis of the contour integral (\ref{generic-eq}) averaged in
this way is still rather involved, but at high intensity, one can
view the spectrum just as a superposition of Gaussians. Inserting
(\ref{Gauss}) into Eq.~(\ref{averaging-positron-def}) then yields
\begin{equation}\label{D-dep-aver-Gauss}
\frac{dw_{e^{+}}}{d\omega}=\frac{\mathcal{D}}{\sqrt{2\pi
\overline{\omega_{1}^2 }(E_{\perp\max})}}\int_0^1
d\xi\xi^{\mathcal{D}-3/2}
\exp\left\{-\frac{\left[\omega-\xi\overline{\omega_{1}
}(E_{\perp\max})\right]^2}{2\xi\overline{\omega_{1}^2
}(E_{\perp\max})}\right\}.
\end{equation}
The latter integral can be expressed in terms of error functions,
but we will restrict ourselves to elucidating its qualitative
behavior.

At $\omega<\overline{\omega_{1} }(E_{\perp\max})-\Delta$, with
$\Delta\gg\sqrt{\overline{\omega_{1}^2 }(E_{\perp\max})}$, the
maximum of the integrand at $\xi\approx
{\omega}/{\overline{\omega_{1} }(E_{\perp\max})}$ falls well within
the integration interval, thereby giving the dominant contribution
to the integral. Neglecting contributions from the endpoints leads
to the result
\begin{equation}\label{dwch-power-asympt}
\frac{dw_{e^{+}}}{d\omega}\underset{\omega<\overline{\omega_{1}
}(E_{\perp\max})-\Delta}\simeq
\frac{\mathcal{D}}{\omega}\left[\frac{\omega}{\overline{\omega_{1}
}(E_{\perp\max})}\right]^{\mathcal{D}}.
\end{equation}
It is independent of the variance $\overline{\omega_{1}^2
}(E_{\perp\max})$, and might be obtained by replacing partial
Gaussians (including their pre-exponential factor) in (\ref{D-dep-aver-Gauss}) by the infinitesimally narrow
distribution
\[
\delta\left[\omega-\xi\overline{\omega_{1}
}(E_{\perp\max})\right].
\]
That also demonstrates the existence of an approximate one-to-one
correspondence between $\omega$ and the transverse energy of the
positron quantified by the ratio $\xi=E_{\perp}/E_{\perp\max}$.

\begin{figure}
\includegraphics[width=85mm]{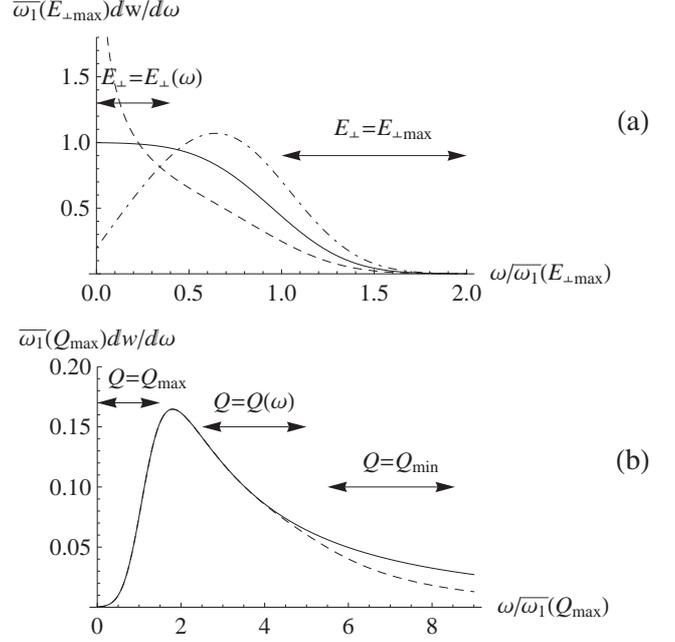}
\caption{\label{fig:aver-Gaussians} (a) High-intensity
approximations (\ref{D-dep-aver-Gauss}) for multiphoton channeling
radiation from positrons, for $\langle n
\rangle\mathcal{D}=n_{\max}\sim \overline{\omega_{1}
}^2(E_{\perp\max})/\overline{\omega_{1}^2 }(E_{\perp\max})=10$.
Dashed curve, planar channeling radiation from monokinetic beam,
$\mathcal{D}=1/2$. Solid curve, planar channeling radiation,
$\mathcal{D}=1$. Dot-dashed curve, axial  channeling radiation,
$\mathcal{D}=2$. (b) High-intensity approximations for multiphoton
channeling radiation from axially channeled electrons. Solid curve,
Eq.~(\ref{e-ax-Gauss-integral}), for $\overline{\omega_{1}
}^2(T_{\max},\Lambda_{\max})/\overline{\omega_{1}^2
}(T_{\max},\Lambda_{\max})=10$. Dashed curve, the same with a lower
cutoff $\lambda_{\min}=\tau_{\min}=0.2$.}
\end{figure}

On the contrary, beyond the break, the integrand exponentially
increases throughout the integration interval. Therefore, the main
contribution to the integral comes from a vicinity of the endpoint
$\xi=1$, around which the $\xi$-dependence of the exponent can be
linearized, whereas $\xi^{\mathcal{D}-3/2}$ be put to unity, giving
\begin{equation}\label{dwch-exp-asympt}
\frac{dw_{e^{+}}}{d\omega}\underset{\omega> \overline{\omega_{1}
}(E_{\perp\max})+\Delta}\simeq
\frac{\mathcal{D}}{\sqrt{\pi}}\frac{\sqrt{2\overline{\omega_{1}^2
}(E_{\perp\max})}}{\omega^2-\overline{\omega_{1}
}^2(E_{\perp\max})}e^{-\frac{\left[\omega-\overline{\omega_{1}
}(E_{\perp\max})\right]^2}{2\overline{\omega_{1}^2
}(E_{\perp\max})}}.
\end{equation}

Examples of behavior of function (\ref{D-dep-aver-Gauss}) are
shown in Fig.~\ref{fig:aver-Gaussians}(a). The spectra strongly
depend on the value of $\mathcal{D}$, especially at low $\omega$, in
accord with asymptotics (\ref{dwch-power-asympt}). That
illustrates the profound effect of the averaging on multiphoton
channeling radiation spectra.



\subsection{Axial channeling of electrons}

Compared to positrons, channeling of electrons, generally, is
affected by dechanneling stronger, because electrons are attracted
to atomic nuclei, and thus may enter regions of strong scattering.
For axial orientation of the crystal, though, electrons can orbit
around positively charged atomic strings, and thus keep on distance
from nuclear vicinities. The approximate axial symmetry of a string
alone, without introducing oversimplified assumptions about the
$r_{\perp}$ dependence, does not yet permit explicitly expressing
electron trajectories and therefrom the radiation spectra.
Nonetheless, qualitative analysis is feasible for a potential
modeled by a pure logarithmic function with a sharp cutoff at large
radial distance $r_{\perp}$:
\begin{equation}\label{V-log}
V(r_{\perp})=V_0\ln\frac{r_{\perp}}{r_{a}}\vartheta(r_{a}-r_{\perp}),
\qquad r_{\perp}< r_{a}.
\end{equation}

The motion of electrons in potential (\ref{V-log}) is determined by two integrals of motion: transverse energy $E_{\perp}$ and projection $\Lambda$ of the angular momentum on the string direction. 
Estimates of first moments of the single-photon spectrum in a crystal of definite thickness give:
\begin{equation}\label{mean-omega-ax}
\overline{\omega_{1}}=\int_0^{\infty}d\omega_1\frac{dI}{d\omega_1}\propto\frac1{\Lambda T},
\end{equation}
and
\begin{equation}\label{mean-omega2-ax}
\overline{\omega^2_{1}}=\int_0^{\infty}d\omega_1\omega_1\frac{dI}{d\omega_1}\sim\frac{4\pi\gamma^2}{T} \overline{\omega_{1}}\propto\frac1{\Lambda T^2},
\end{equation}
where $T$ is the period of the radial motion, exponentially
depending on $E_{\perp}/V_0$, and relatively weakly dependent on
$\Lambda$. Both spectral moments blow up as $T\to0$ or
$\Lambda\to0$, i.e., when the electron passes close to the
singularity of the continuous potential.

The element of the uniformly populated phase space can be expressed
in terms of $T$, $\Lambda$, as well:
\begin{equation}\label{phase-space-el}
d^2r_{\perp}d^2p_{\perp}\propto T dE_{\perp}d\Lambda\approx V_0 dT
d\Lambda,
\end{equation}
and the region accessible to channeling (subject to condition
$E_{\perp}<0$)\footnote{Here we refrain ourselves from consideration
of channeled particles with $E_{\perp}>0$, which exist for axial
channeling. The contribution from such particles is relatively small
and will be analyzed elsewhere.} is
\[
0\leq\lambda=\frac{|\Lambda|}{\Lambda_{\max}}\leq\tau=\frac{T}{T_{\max}}\leq1,
\]
where $T_{\max}\approx 3r_{a}\sqrt{E_e/V_0}$,
$\Lambda_{\max}=r_{a}\sqrt{E_eV_0/e}$. Ultimately, the averaged
multiphoton spectrum assumes the form
\begin{eqnarray}\label{e-ax-Gauss-integral}
\frac{dw_{e^-\text{ax}}}{d\omega}=\frac{2}{\sqrt{2\pi\overline{\omega_{1}^2
}(Q_{\max})}}\qquad\qquad\qquad\qquad\qquad\quad\nonumber\\
\times\int_0^1 d\tau\tau \int_0^{\tau}d\lambda\sqrt{\lambda}
\exp\left\{-\frac{\left[\omega\lambda\tau-\overline{\omega_{1}
}(Q_{\max})\right]^2}{2\lambda\overline{\omega_{1}^2
}(Q_{\max})}\right\},
\end{eqnarray}
with $Q_{\max}=T_{\max}\Lambda_{\max}$. Again, we restrict ourselves
to qualitative analysis of integral (\ref{e-ax-Gauss-integral}).

At $\omega<\overline{\omega_{1} }(Q_{\max})-\Delta$,
$\Delta\gg\sqrt{\overline{\omega_{1}^2}(Q_{\max})}$, the maximum of
the integrand at $\lambda\tau={\overline{\omega_{1}
}(Q_{\max})}/{\omega}$ is beyond the integration domain. Therefore,
the main contribution comes from the endpoint $\lambda=\tau=1$, and
generally, the integral is exponentially small:
\begin{eqnarray}\label{}
\frac{dw_{e^-\text{ax}}}{d\omega}\!\!\underset{\omega<
\overline{\omega_{1}
}(Q_{\max})-\Delta}\simeq\!\!\sqrt{\frac2{\pi}}\frac{2\overline{\omega_{1}^2
}^{3/2}(Q_{\max})}{\left[\overline{\omega_{1}
}^2(Q_{\max})-\omega^2\right]\left[\overline{\omega_{1}
}(Q_{\max})+\omega\right]}\nonumber\\
\times\exp\left\{-\frac{\left[\omega-\overline{\omega_{1}
}(Q_{\max})\right]^2}{2\overline{\omega_{1}^2
}(Q_{\max})}\right\}.\qquad\quad
\end{eqnarray}

On the contrary, at $\omega>\overline{\omega_{1}
}(Q_{\max})+\Delta$, the maximum of the integrand falls well within
the integration domain. Neglecting the endpoint contributions, or
simply inserting
\begin{equation}\label{del}
\delta\left[\omega-\overline{\omega_{1}
}(Q_{\max})/\lambda\tau\right]
\end{equation}
in place of the partial Gaussians, the asymptotics of the
multiphoton spectrum derives as
\begin{equation}\label{omega-2log}
\frac{dw_{e^-\text{ax}}}{d\omega}\underset{\omega>
\overline{\omega_{1}
}(Q_{\max})+\Delta}\simeq\frac{\overline{\omega_{1}
}(Q_{\max})}{\omega^2}\ln\frac{\omega}{\overline{\omega_{1}
}(Q_{\max})}.
\end{equation}
The corresponding mean energy loss $\int_0^{\infty}d\omega\omega \frac{dw_{e^-\text{ax}}}{d\omega}$ diverges in the ultraviolet as a double logarithm, in accordance with the double logarithmic divergence of the single-photon average mean energy loss, cf. Eq.~(\ref{mean-omega-ax}).

The delta function (\ref{del}) implies that a correspondence between
$\omega$ and the integrals of motion holds for electron axial
channeling radiation, as well. But in contrast to radiation from
channeled positrons,  the correspondence here takes place at the
outer slope of the spectrum [see Fig.~\ref{fig:aver-Gaussians}(b)].
Yet, since there are two integrals of motion, the correspondence of
$\omega$ with them is no longer one-to-one, unless one refers to the
special integral of motion $Q=T\Lambda\propto
e^{E_{\perp}/V_0}\Lambda$.

Experimentally measured multiphoton axial channeling radiation
spectra \cite{Bavizhev,Avakian} decrease considerably faster than by
power law (\ref{omega-2log}). That must be attributed to the
crudeness of our model (\ref{V-log}) and the physical finiteness of
$\left\langle \overline{\omega_{1} } \right\rangle$ due to the
smearing of the string potential at low $r_{\perp}$ through thermal
and quantum vibrations. The simplest way to take all that into
account is to introduce a lower cutoff $\lambda_{\min}=\tau_{\min}$
into integrals (\ref{e-ax-Gauss-integral}), which leads to an
exponentially decreasing spectrum [see
Fig.~\ref{fig:aver-Gaussians}(b), dashed curve]. The beginning of the exponential suppression region depends on the value of the
cutoff parameter and therethrough on the crystal temperature.

\section{Limitations and challenges}

The intrinsic simplicity of the soft photon resummation procedure
makes it a useful tool in various physical situations, predicting
generation of higher quasi-harmonics at moderate intensity (as discussed
in \cite{Bondarenco-paper}) and the existence of a set of limiting
distributions at high intensity, depending on the shape of the
confining well, yet yielding in certain spectral regions approximate
correspondence between the radiative loss $\omega$ and one of the
integrals of motion of the channeled particle.

At the same time, the resummation method has its limitations and is
unable to cover, e.g., the case when a considerable part of the
spectrum concentrates near the kinematic edge $\omega=E_e$
\cite{Belkacem}. Also, even in the domain of applicability of the
soft photon approximation (\ref{omega-ll-Ee}), one may wish to
evaluate higher order corrections in $\omega/E_e$, e.g., in order to
account for degradation of $E_e$ stipulating additional broadening
and redshift of the multiphoton spectra \cite{Potyl-Chann2014}, or
to describe the effect of electron beam bunching in FELs
\cite{FEL-multiphot}.

For the specific problem of channeling radiation, there arise
additional issues. First is the need for knowledge of not only the
averaged single-photon spectrum, which might be measured at a
sufficiently small crystal thickness, but also of the dependence of
the radiation spectrum on electron entrance parameters. That can be
illustrated, e.g., by the variance of the multiphoton spectrum about
the mean value, for positron channeling in a harmonic well discussed
in Sec.~\ref{subsec:positron-ch}:
\begin{equation}\label{25}
\left\langle\overline{
(\omega-\left\langle\overline\omega\right\rangle)^2}\right\rangle=\left\langle\overline{\omega^2_1}\right\rangle+\left\langle\overline{\omega_1}\right\rangle^2\frac{1}{\mathcal{D}(\mathcal{D}+2)}.
\end{equation}
Due to the second term, it depends on the target thickness
\emph{nonlinearly}, yet involving a dependence on the value of
$\mathcal{D}$, which in turn depends on the shape of the well. Also,
the existence of a correspondence in some spectral regions between
$\omega$ and the particle integrals of motion implies that not all
the values of integrals of motion contribute at a given $\omega$, in
contrast to the case of the averaged single-photon spectrum.

Further complications emerge because in the presence of
dechanneling, relative contributions of different integrals of
motion can attenuate differently. The corresponding factors thus
must be introduced into integrals (\ref{D-dep-aver-Gauss}),
(\ref{e-ax-Gauss-integral}). Thirdly, the dechanneled particles
contribute to the radiation spectrum, as well, although in a
somewhat harder spectral region. That demands inclusion of their
contribution to the multiphoton spectrum, too. On the other hand,
studying multiphoton radiation spectra under such conditions may
offer deeper insight into the complex dynamics of channeling.







\begin{thebibliography}{00}


\bibitem{Bak}
J.~Bak \emph{et al.}, Nucl. Phys. B \textbf{254} (1985) 491; \textit{ibid}. \textbf{302} (1988) 525.

\bibitem{Telnov}
V.~Telnov, NIM A \textbf{355} (1995) 3.

\bibitem{DiPiazza}
A.~Di Piazza, C.~M\"{u}ller, K.Z.~Hatsagortsyan, and C.H.~Keitel.
Rev. Mod. Phys. \textbf{84} (2012) 1177; A.~Di Piazza,
K.Z.~Hatsagortsyan, and C.H.~Keitel. Phys. Rev. Lett. \textbf{105}
(2010) 220403.

\bibitem{AccelHandbook}

A.W. Chao and M. Tigner (eds.), \textit{Handbook of Accelerator Physics and Engineering}, World Scientific, Singapore, 2002.

\bibitem{FEL-multiphot}

A.~Friedman \emph{et al.}, Rev. Mod. Phys. \textbf{60} (1988) 471.





\bibitem{Glauber}
R.J.~Glauber, Phys. Rev. \textbf{84} (1963) 395;
A.I.~Akhiezer and V.B.~Berestetskii, \textit{Quantum Electrodynamics}, Nauka, Moscow, 1981 (in Russian).

\bibitem{BK}
V.N.~Baier and V.M.~Katkov, Phys. Rev. D \textbf{59} (1999) 056003.


\bibitem{Bondarenco-paper}
M.V.~Bondarenco, Phys. Rev. D \textbf{90} (2014) 013019.


\bibitem{Bondarenco-RREPS2013}

M.V.~Bondarenco, J. Phys. Conf. Ser. \textbf{517} (2014) 012027.



\bibitem{EW-boson-resum}
G.~Parisi and R.~Petronzio, Nucl. Phys. B \textbf{154} (1979) 427;

J.~Collins, D. Soper, and G.~Sterman, Nucl. Phys. B \textbf{250}
(1985) 199.

\bibitem{BKS-UFN}
V.N.~Baier, V.M.~Katkov, and V.M.~Strakhovenko,  Sov. Phys. Usp. \textbf{32} (1989) 972.

\bibitem{Artru-code}

X.~Artru, NIM B \textbf{48} (1990) 278.



\bibitem{Guidi-Bandiera-Tikhomirov}
V.~Guidi, L.~Bandiera and V.~Tikhomirov, Phys. Rev. A \textbf{86} (2012) 042903.













\bibitem{Lietti}
D.~Lietti \emph{et al.}, NIM B \textbf{283} (2012) 84.

\bibitem{Bond-CBBC}
M.V.~Bondarenco, Phys. Rev. A \textbf{81} (2010) 052903.

\bibitem{Bondarenco-RREPS2010}

M.V.~Bondarenco, J. Phys. Conf. Ser. \textbf{236} (2010) 012026.



\bibitem{Bavizhev}
M.D.~Bavizhev, Yu.V.~Nil'sen, and B.A.~Yur'ev, Zh. Eksp. Teor. Fiz.
\textbf{95} (1989) 1392 [Sov. Phys. JETP \textbf{68} (1989) 803].

\bibitem{Avakian}
R.~Avakian \emph{et al.}, Sov. Tech. Phys. Lett. B \textbf{14} (1988) 395.




\bibitem{Belkacem}
A.~Belkacem \emph{et al.}, Phys. Rev. Lett. \textbf{54} (1985) 2667;
Phys. Lett. B \textbf{177} (1986) 211; Europhys. Lett. \textbf{5}
(1988) 589.



\bibitem{Potyl-Chann2014}

A.P.~Potylitsin and A.M.~Kolchuzhkin, Phys. Part. Nucl. \textbf{45} (2014) 1000.



\end{thebibliography}
\end{document}